# Microresonator Kerr frequency combs with high conversion efficiency

Xiaoxiao Xue,[1,2,*] Pei-Hsun Wang,[2] Yi Xuan,[2,3] Minghao Qi,[2,3] and Andrew M. Weiner[2,3]

[1]Department of Electronic Engineering, Tsinghua University, Beijing 100084, China
[2]School of Electrical and Computer Engineering, Purdue University, 465 Northwestern Avenue, West Lafayette, Indiana 47907-2035, USA
[3]Birck Nanotechnology Center, Purdue University, 1205 West State Street, West Lafayette, Indiana 47907, USA
*Corresponding author: xuexx@tsinghua.edu.cn

**ABSTRACT:** Microresonator-based Kerr frequency comb (microcomb) generation can potentially revolutionize a variety of applications ranging from telecommunications to optical frequency synthesis. However, phase-locked microcombs have generally had low conversion efficiency limited to a few percent. Here we report experimental results that achieve ~30% conversion efficiency (~200 mW on-chip comb power excluding the pump) in the fiber telecommunication band with broadband mode-locked dark-pulse combs. We present a general analysis on the efficiency which is applicable to any phase-locked microcomb state. The effective coupling condition for the pump as well as the duty cycle of localized time-domain structures play a key role in determining the conversion efficiency. Our observation of high efficiency comb states is relevant for applications such as optical communications which require high power per comb line.

Microresonator-based optical Kerr frequency comb (microcomb) generation is a very promising technique for portable applications due to its potential advantages of low power consumption and chip-level integration [1]. In the past decade, intense researches have been dedicated to investigating the mode-locking mechanism [2-9], dispersion and mode engineering [10-16], searching for new microresonator platforms [17-24], and reducing the microresonator losses. Very low pump power in the milliwatt level has been achieved by using microresonators with high quality factors [17, 18, 24-26]. Another important figure of merit is the power conversion efficiency, i.e. how much power is converted from the single-frequency pump to the generated new frequency lines. Most phase coherent microcombs have poor conversion efficiency, which is generally indicated by a large contrast between the residual pump power and the power level of the other comb lines in the waveguide coupled to the microresonator [4-7, 9]. The conversion efficiency is particularly important for applications that employ each comb line as an individual carrier to process electrical signals, such as fiber telecommunications [27, 28] and radiofrequency (RF) photonic filters [29, 30]. In those systems, the power level of each comb line usually plays a key role in determining the overall electrical-to-electrical noise figure and insertion loss. An analytical and numerical analysis on the efficiency of bright soliton combs in the anomalous dispersion region was reported in [31]. It was shown that the conversion efficiency of bright solitons is generally limited to a few percent, a finding confirmed by experiments such as [35]. In this letter, we report experimental results of mode-locked microcombs with much higher conversion efficiency (even exceeding 30%) in the fiber telecom band by employing dark pulse mode-locking in the normal dispersion region.

We begin by presenting a general analysis which provides useful insights into the efficiency of any comb state that is phase-locked. Figure 1 shows the energy flow chart in microcomb generation. Considering the optical field circulating in the cavity, part of the pump and comb power is absorbed (or scattered) due to the intrinsic cavity loss, and part is coupled out of the cavity into the waveguide; meanwhile, a fraction of the pump power is converted to the other comb lines, effectively resulting in a nonlinear loss to the pump. At the output side of the waveguide, the residual pump line is the coherent summation of the pump component coupled from the cavity and the directly transmitted pump; the power present in the other comb lines excluding the pump constitutes the usable comb power. Here we omit any other nonlinear losses possibly due to Raman scattering or harmonic generation. Energy equilibrium is achieved when the comb gets to a stable phase-locked state.

The power conversion efficiency is defined as

$$\eta = P_{\text{other}}^{\text{out}} \big/ P_{\text{pump}}^{\text{in}} \qquad (1)$$

where $P_{\text{pump}}^{\text{in}}$ is the input pump power and $P_{\text{other}}^{\text{out}}$ is the power of the other comb lines (i.e., excluding the pump) at the waveguide output. Note that 100% conversion efficiency requires complete depletion of the pump at the output. Indeed, one important factor related to the conversion efficiency is the reduction in pump power after the microresonator compared to the input. A significant pump reduction does not necessarily mean, but is a prerequisite to, significant conversion efficiency. By considering both the linear and effective nonlinear losses to the pump, the effective complex amplitude transmission for the pump line is given by (see the supplementary information of [8])

$$T_{\text{eff}} = E_{\text{pump}}^{\text{out}} \big/ E_{\text{pump}}^{\text{in}} = 1 - \theta \big/ (\alpha_{\text{eff}} + i\delta_{\text{eff}}) \qquad (2)$$

where $\theta$ is the waveguide-cavity power coupling ratio, $\alpha_{\text{eff}}$ is the total effective cavity loss for the pump, and $\delta_{\text{eff}}$ is the effective pump-resonance phase detuning under comb operation. Note that all the lost energy originates from the pump, thus the nonlinear loss for the pump due to comb generation should equal the total cavity loss experienced by the other comb lines, i.e. $\alpha_{\text{eff}} = \alpha \cdot P_{\text{all}}^{\text{cavity}} \big/ P_{\text{pump}}^{\text{cavity}}$ where $\alpha$ is the total cavity loss including the intrinsic loss and the coupling loss, $P_{\text{all}}^{\text{cavity}}$ is the total optical power in the cavity, $P_{\text{pump}}^{\text{cavity}}$ is the pump power in the cavity. Here we assume the linear cavity loss is uniform for all the frequencies. For ultra-broadband combs spanning nearly one octave, the cavity loss may vary with the comb lines due to the frequency dependences of the material absorption loss, the scattering loss, and the coupling loss. In this case, the effective loss for the pump should be modified to $\alpha_{\text{eff}} = \sum_l \alpha(\omega_l) P(\omega_l) \big/ P_{\text{pump}}^{\text{cavity}}$ where $P(\omega_l)$ is power of the $l$ th comb line and $\alpha(\omega_l)$ is the corresponding cavity loss.

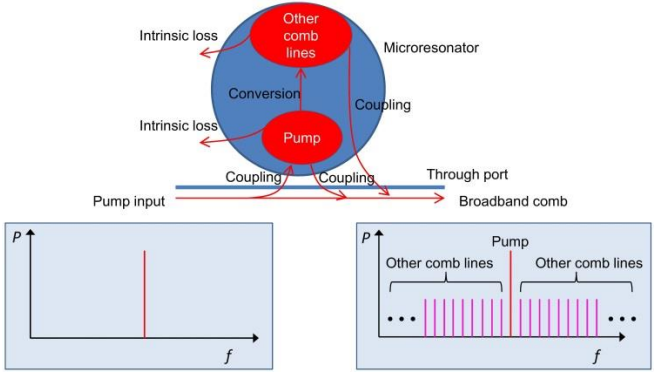

Fig. 1. Energy flow chart in microcomb generation.

A large reduction in pump power can be achieved when the cavity under comb operation is effectively critically coupled for the pump and when the effective phase detuning is close to zero. To achieve effective critical coupling requires $\alpha_{\text{eff}} = \theta$, i.e. $\alpha(1+k) = \theta$ where $k = P_{\text{other}}^{\text{cavity}}/P_{\text{pump}}^{\text{cavity}}$ is the power ratio in the cavity of the other comb lines (excluding the pump) to the pump. Note that the total linear field amplitude loss in the cavity is given by $\alpha = (\alpha_i + \theta)/2$ where $\alpha_i$ is the intrinsic power loss [33]. The conversion efficiency can in general be improved by increasing the ratio of $\theta$ to $\alpha_i$ as has been proposed in literature [32, 34], since it reduces the power fraction lost due to the intrinsic loss. Here we consider the ideal case of a cavity which has zero intrinsic loss ($\alpha_i = 0$, corresponding to an infinite intrinsic Q). Practically this will be approximately true when the microresonator is strongly over-coupled. Taking $\alpha = \theta/2$ and setting $\alpha_{\text{eff}} = \alpha(1+k) = \theta$ for effective critical coupling, we find $k = 1$. This means that for effective critical coupling, the total power of the generated new comb lines in the cavity should equal that of the intracavity pump.

The effective pump-resonance phase detuning in comb operation depends on the specific mode-locking mechanism. We numerically simulated two broadband comb states that are widely investigated in the literature – bright solitons in the anomalous dispersion region and dark pulses in the normal dispersion region. The simulations are based on the mean-field Lugiato–Lefever equation [33]. The effective cavity loss and detuning are retrieved based on Eq. (2) after the intracavity optical field evolves to a stable state. Typical parameter values for silicon nitride microrings are used except that the intrinsic loss is assumed zero; $\alpha_i = 0$, $\theta = 1.22 \times 10^{-2}$ (corresponding to a loaded Q of $1 \times 10^6$), $FSR = 100$ GHz (equal to the standard channel spacing defined by ITU-T for wavelength-division multiplexing fiber telecommunications), $P_{\text{pump}}^{\text{in}} = 600$ mW, $\delta_0 = 0.1$ rad (cold-cavity detuning), $\gamma = 1$ m$^{-1}$W$^{-1}$ (nonlinear Kerr coefficient), $\beta_2 = -200$ ps$^2$/km for anomalous dispersion and 200 ps$^2$/km for normal dispersion.

Figure 2(a) shows the spectra for the bright-soliton comb. The power ratio of the comb lines excluding the pump is 69.2% in the cavity, corresponding to relatively high internal conversion efficiency. However, in the waveguide the power ratio of the generated new comb lines is only 2.5%. The pump power at the waveguide output drops by only 0.1 dB compared to the input. The large contrast between the internal and external conversion efficiencies is consistent with the experimental observations in [35]. Figure 2(c) shows the effective cavity transmission and the effective detuning for the pump corresponding to Eq. (2). Note that the microresonator is assumed intrinsically lossless in the simulations, thus the cold-cavity transmission is all-pass. However, due to the loss to the pump caused by power transfer to the comb, the effective hot-cavity transmission shows a dip with a moderate extinction ratio of ~8 dB. Nevertheless, the effective pump detuning is −1.8 GHz (the corresponding retrieved $\delta_{\text{eff}}$ is 0.11 rad; the relation between the frequency and phase detunings is given by $\Delta f_{\text{eff}} = \delta_{\text{eff}}/(2\pi) \cdot FSR$), which is much larger than the effective resonance width (628 MHz, calculated by $B = \alpha_{\text{eff}} \cdot FSR/\pi$ where $\alpha_{\text{eff}} = 1.97 \times 10^{-2}$) in magnitude. The large effective detuning prevents efficient injection of the pump power into the cavity, and therefore limits the external conversion efficiency.

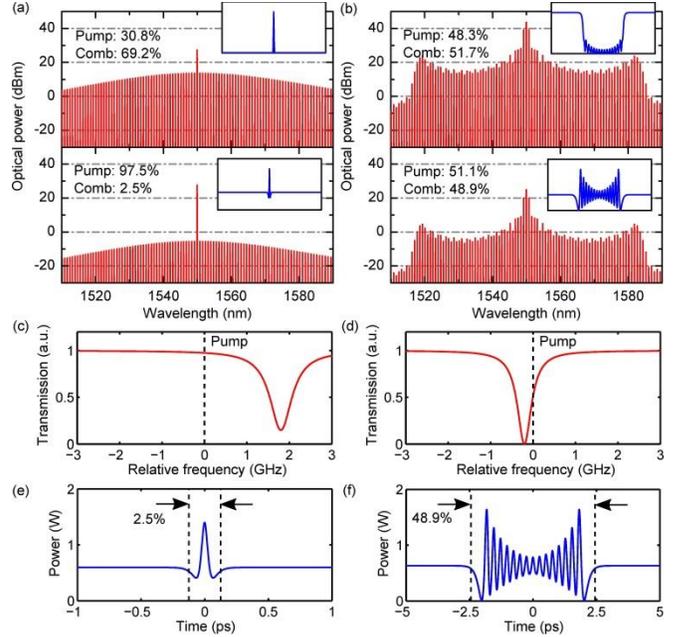

Fig. 2. Comparison of comb generation in anomalous (a, c, e) and normal (b, d, f) dispersion regions (see main text for the simulation parameters). (a, b) Comb spectra. Upper: intracavity, lower: in the waveguide. The power ratios of the pump and the other comb lines are listed in the figure. The insets show the time-domain waveforms. (c, d) Hot-cavity resonance and effective detuning in comb operation. (e, f) Zoom-in waveforms at the waveguide output, showing the relation between time-domain features and conversion efficiency.

Figure 2(b) shows the spectrum for the dark-pulse comb. The power ratio of the comb lines excluding the pump is 51.7% in the cavity, and is 48.9% in the waveguide. The pump power at the waveguide output drops by 2.9 dB compared to the input. As indicated by the insets, the time-domain waveform is a dark pulse inside the cavity, but sits on top of the transmitted pump outside the cavity. Figure 2(d) shows the effective cavity transmission and the effective detuning for the pump. Since the power of the generated new comb lines roughly equals that of the pump in the cavity, the resonance dip gets very close to the critical coupling condition. The effective pump detuning is 204 MHz which is smaller than the effective resonance width (400 MHz). Compared to the bright-soliton case, the smaller effective detuning here is one important reason that the dark-pulse comb can achieve higher external conversion efficiency. The advantageous conversion efficiency of dark-pulse combs was also demonstrated numerically in [36]. It should be noted that the dark-pulse comb may show larger spectral modulation compared to a single bright soliton comb. In Fig. 2(b), the two comb lines adjacent to the pump are much stronger than the other comb lines.

The time-domain waveforms also provide useful insights into the conversion efficiency. Both bright and dark pulses are localized structures sitting on a pedestal. In the case of intrinsically lossless cavities, the pedestal level is exactly equal to the input pump power (recall that intrinsically lossless cavities have all-pass transmission). Furthermore, the instantaneous frequency of the pedestal is also equal to that of the pump. In other words, the pump energy is completely unconverted over much of the pedestal. The energy of the converted new frequency lines concentrates in the time region where the solitary structure is located. From this point of view, the duty cycle of the solitary wave provides a good estimation on the conversion efficiency, i.e. $\eta \sim \Delta T/t_R$ where $\Delta T$ is the time width of the localized structure and $t_R$ is the round trip time. Figures 2(e) and 2(f) show the zoom-ins of the output waveforms for anomalous and normal dispersion, respectively. Also shown as a guide are dashed lines indicating a duty cycle exactly equal to the conversion efficiency. A close match to the durations of the localized structures can be observed. Therefore, a time-domain explanation of the higher conversion efficiency predicted for dark-pulse mode-locking is that the dark pulses can be much wider than the bright solitons under similar conditions (dispersion magnitude, Kerr coefficient, pump level, and microresonator Q factor). Another useful conclusion is that the conversion efficiency of bright-soliton combs will be lower if the comb bandwidth gets larger since larger spectral bandwidth usually corresponds to narrower pulse in the time domain. In comparison, dark-pulse combs are largely free from such degradation because more bandwidth can be achieved with sharper rise and fall times while the duty factor is kept the same. This difference has been observed in numerical simulations in [31] and [36]. The efficiency of bright-soliton combs can be increased by increasing the number of solitons in the cavity thus increasing the overall duty cycle. This mechanism provides one possible explanation for recently reported high-efficiency mid-IR comb generation in silicon microrings [37]. However, multiple-bright-soliton combs generally exhibit random soliton number and positions in mode-locking transition [6]. Thus multiple soliton states usually have poor repeatability. In some cases, the soliton positions may be regulated by mode crossings, giving rise to soliton crystals [38]. But the comb power typically concentrates in few lines spaced by multiple FSRs while most other 1-FSR spaced lines are very weak. Such combs may be poorly matched to certain applications, such as optical communications, in which large power variation between optical carriers is undesirable.

Figure 3(a) shows an experimental dark-pulse comb from a normal-dispersion silicon nitride microring (ring 1) measured with 10 GHz spectral resolution. Similar spectra generated with the same ring were shown in our previous report on mode-locked dark pulses [8]. The microring has a radius of 100 μm corresponding to an FSR of ~231 GHz, and a loaded Q factor of $7.7 \times 10^5$. The microring is over-coupled, and the extinction ratio of the cold-cavity transmission is around 4.7 dB. The on-chip pump power is 656 mW at the waveguide input and drops by 4.5 dB after the microring. Figure 3(b) shows the energy flow chart. The external conversion efficiency is 31.8%, corresponding to an on-chip comb power of 209 mW excluding the pump. There are 40 lines including the pump in the wavelength range from 1513 nm – 1586 nm. The average power per comb line excluding the pump is 7 dBm. The strongest line is 17 dBm while the weakest is -6 dBm (in comparison, the residual pump is 23.7 dBm). Much of the comb spectrum sits on top of an ASE pedestal from the amplified pump laser, which could be eliminated by optical filtering prior to the microring. However, even with this ASE, for a significant number of comb lines, the optical signal-to-noise ratio (OSNR) exceeds 40 dB (in some cases 50 dB). Figure 3(c) shows the transmission curves of the cold and hot cavities for the pump. The effective hot cavity detuning is retrieved based on Eq. (2); and the cold cavity detuning is obtained through numerical simulations which mimic the experimental observations (see the simulations presented in [8]). Note that the hot cavity gets closer to critical coupling in comb operation. Such frequency comb enhanced coupling was also observed in our previous experiments [32], where the fraction of pump power that is transmitted past an initially over-coupled microcavity drops by >10 dB above the comb threshold. The effective pump detuning is 119 MHz (in comparison, the effective resonance width is 336 MHz). The inset of Fig. 3(a) shows the time-domain waveform at the waveguide output measured with self-referenced cross-correlation [8]. Since the microring here has an intrinsic loss, the relation between the duty cycle and the conversion efficiency is modified as $\eta \sim \Delta T/t_R \cdot \theta/(\theta+\alpha_i)$ where $\theta/(\theta+\alpha_i)$ represents the proportion of power coupled out of the cavity. The duty cycle retrieved from the conversion efficiency is also plotted in the inset of Fig. 3(a), which is close to the actual width of the localized structure.

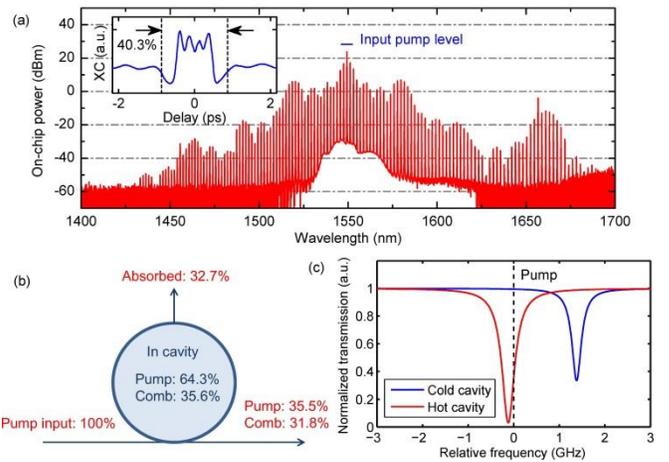

Fig. 3. Experimental results of a mode-locked dark-pulse comb from a microresonator coupled to a bus waveguide (ring 1). (a) Spectrum measured at the waveguide output. The inset shows the time-domain waveforms measured with self-referenced cross-correlation (XC). (b) Energy flow chart. (c) Cold/hot resonances and the pump detuning.

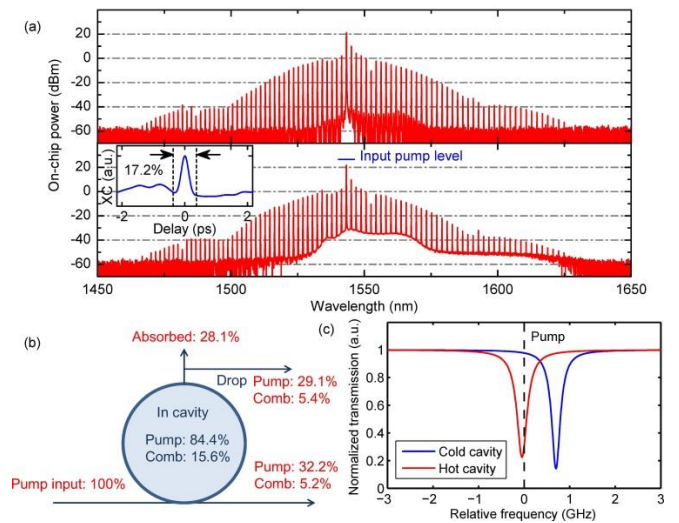

Fig. 4. Experimental results of a mode-locked dark-pulse comb from a microresonator with both a through port and a drop port (ring 2). (a) Spectra measured at the drop port (upper) and the through port (lower). The inset shows the time-domain waveforms at the through port, measured with self-referenced cross-correlation (XC). (b) Energy flow chart. (c) Cold/hot resonances and the pump detuning.

Figure 4(a) shows another experimental example of a dark-pulse comb from a different microring (ring 2). This microring has both a through port and a drop port with symmetric coupling gaps, resulting in an under-coupling condition. The loaded Q is $8.6 \times 10^5$; the resonance dip is around 8.5 dB. The same ring was also used in our previous report [8]; the comb presented in the current Letter is generated by exploiting a different resonance [39]. The on-chip pump power is 454 mW at the waveguide input and drops by 5 dB after the microring. Figure 4(b) shows the energy flow chart. The conversion efficiency at the through port is 5.2%, corresponding to an on-chip comb power of 24 mW excluding the pump. A frequency comb with similar power level is also obtained at the drop port. Thus the overall conversion efficiency is 10.6%. The inset of Fig. 4(a) shows the time-domain waveform at the through port measured with self-referenced cross-correlation [8]. Here the relation between the soliton duty cycle and the overall efficiency is given by $\eta \sim \Delta T/t_R \cdot 2\theta/(2\theta+\alpha_i)$. Again the duty cycle retrieved from the conversion efficiency is close to the actual width of the localized structure. Figure 4(c) shows the transmission curves of the cold and hot cavities for the pump. The hot cavity gets further under-coupled in comb operation. The effective pump detuning is 46 MHz (in comparison, the effective resonance width is 270 MHz).

We note that the higher conversion efficiency achieved with ring 1 compared to ring 2 is due to two reasons. First, ring 1 is over-coupled while ring 2 is under-coupled. In comb generation, ring 1 gets closer to critical coupling for the pump, thus facilitating efficient pump injection into the cavity. Actually ring 1 has a larger effective pump detuning in Fig. 3 compared to ring 2 in Fig. 4. But the pump power drop through the microring is similar in both cases (~5 dB). Second, the duty cycle of the dark pulse in ring 1 is larger than that in ring 2, corresponding to a larger fraction of pump power converted to new frequencies.

In summary, we have demonstrated microcomb generation with a high conversion efficiency up to 31.8% by employing dark-pulse mode-locking in the normal dispersion region. This corresponds to 209 mW on-chip comb power excluding the pump. The high efficiency of dark-pulse combs makes them good candidates for fiber telecommunications and RF photonic filtering. A general analysis on the conversion efficiency of microcombs is presented. The effective coupling condition for the pump as well as the duty cycle of localized time-domain structures play a key role in determining the conversion efficiency. Our findings can provide useful guidance for exploiting new high-efficiency microcomb states.

**Funding.** National Science Foundation (ECCS-1509578); Air Force Office of Scientific Research (FA9550-15-1-0211); DARPA PULSE program (W31P40-13-1-0018). X. Xue was supported in part by the National Natural Science Foundation of China (61420106003).

**Acknowledgment.** We thank Dr. Victor Torres-Company for fruitful discussions, and Dr. Daniel E. Leaird and Mr. Jose A. Jaramillo-Villegas for technical help in experiments.